# Impact of UC$_2$, UC, UBC and UB$_2$ target compositions on the release of fission products

Julien Guillot[a*], Brigitte Roussière[a], Sandrine Tusseau-Nenez[b], Isabelle Deloncle[a], Martin Humenny[a], Matthieu Lebois[ad], Jean-François Ledu[a], Jérome Roques[a], François Brisset[c]

[a] Université Paris-Saclay, CNRS/IN2P3, IJCLab, 91405 Orsay, France
[b] Laboratoire de Physique de la Matière Condensée, CNRS, École Polytechnique, Institut Polytechnique de Paris, 91128 Palaiseau, France
[c] Université Paris-Saclay, CNRS, ICMMO, 91405 Orsay, France
[d] Institut Universitaire de France, 1 rue Descartes, 75005 Paris, France

*Corresponding author: julien.guillot@ijclab.in2p3.fr, tel: +33169157248

___

**Highlights**

- Synthesis and characterization of uranium compounds (UC$_2$, UC, UBC, and UB$_2$).
- Measurement the release fractions of 11 fission products, revealing how crystallic configurations and packing fractions influence the release behaviour of different elements.
- Correlated physicochemical properties with release fractions using Principal Component Analysis (PCA).

**Abstract**:

The release properties of 4 targets (UC$_2$, UC, UBC, UB$_2$) were measured for 11 elements (Kr, Sr, Ru, Sn, Sb, Te, I, Cs, Ba, La, and Ce) using an off-line technique. The crystal packing fraction and the size of the studied element play a key role in the release process. However, physicochemical properties are also involved, notably melting and boiling points in vacuum and the minimal oxidation state. Principal component analysis was used to investigate the interrelationships between the physicochemical properties of fission products (from Fe to Dy) and the observed releases, thereby enabling predictions to be made about the release properties of the four crystallic configurations for elements that are inaccessible in off-line experiments.

___

1. Introduction

In the field of nuclear physics, the Isotope Separation Online (ISOL) technique represents one of the principal methods used to produce radioactive ion beams. This process entails the generation of radioactive nuclei through the irradiation of a target at a constant and high temperature, with the objective of rapidly extracting radioactive atoms. This is followed by the ionisation and separation of the desired isotopes. The intensity of the produced radioactive beam ($I$) depends on several factors: the intensity of the primary beam ($I_p$), the quantity of target nuclei ($N$), the production cross-section ($\sigma$) of the nucleus of interest, and the efficiencies of release ($\varepsilon_{release}$), ionisation ($\varepsilon_{ionisation}$), and transport ($\varepsilon_{transport}$). It is expressed as:

$$I = I_p . N . \sigma . \varepsilon_{release} . \varepsilon_{ionisation} . \varepsilon_{transport} \quad (1)$$

The release efficiency of radioactive atoms depends on the diffusion properties of the element in the target material ($\varepsilon_{diff}$) [1], and the atom's effusion through the porosities to the ion source ($\varepsilon_{eff}$) [2]. The overall release efficiency is particularly influenced by the half-life of the isotope, as short-lived isotopes may decay before they are released from the target material. In the case of the spherical grains, the efficiency of release by diffusion is given by:

$$\varepsilon_{diff} = \frac{6}{\pi^2} \sum_{m=1}^{\infty} \frac{1}{m^2 + \frac{\lambda . a^2}{D.\pi^2}} \quad (2)$$

where $D$ is the diffusion coefficient, $\lambda$ represents the radioactive decay constant, $a$ is the radius of the spherical grain. The expression for the effusion efficiency for a radioactive particle can be written as:

$$\varepsilon_{eff} = \frac{1}{1+\lambda.n(t_{fl}+t_{st})} \quad (3)$$

where, $n$ is the number of effusion steps required for the particle to reach the ion source, $t_{fl}$ and $t_{st}$ are respectively the flying time between two collisions and the sticking time of the atom.

A review published in 2019 [3], summarising research efforts to improve this release efficiency, highlighted the importance of acquiring a better understanding of the release mechanisms specific to each element. Differences in release times measured for Cs, Kr, I, and Sn atoms in $UC_x$ targets revealed that while Cs and Kr mainly undergo a diffusion process in the target, Sn release is characterised by an effusive mechanism [4], [5], [6], whereas iodine does not present a dominant release mechanism. These observations demonstrate the interest in adapting the target microstructure to optimise its release properties. In order to favour the diffusion release of short-lived isotopes of elements such as Cs or Kr, it is preferable to use low-density $UC_x$ targets with a porous structure and nanometric grains [7]. Conversely, denser $UC_x$ targets may be advantageous for elements such as Sn.

The work of Kronenberg *et al.* (2008) examined the production of Ga, Br, Sb, and Xe isotopes using $UC_2$ targets of different densities and $UB_4$ and $ThO_2$ targets. The study revealed significant differences in production yields between these materials [8]. It was found that $ThO_2$ and $UB_4$ targets consistently induced yields an order of magnitude lower than those obtained with $UC_2$ targets. This demonstrated the impact of target composition and density on production efficiency. However, the lack of detailed physicochemical characterisation of the targets prevented the extraction of parameters that could explain these differences.

Consequently, our study specifically aims to qualify the influence of different target compositions ($UC_2$, UC, UBC, $UB_2$) on the release of various elements (Kr, Sr, Ru, Sn, Sb, Te, I, Cs, Ba, La, Ce) in an off-line experiment (presented in deeper details in [9]). The targets are composed of sintered polycrystalline grains. The main goal is to identify the physicochemical parameters that may influence the release of the studied radioactive elements in order to predict which elements, not accessible in our study, are likely to be released by one and/or the other of the four studied target materials when operated on-line. The following section describes the different steps for producing the aforementioned elements in an off-line experiment.

2. Experimental part
2.1. Synthesis of $UC_2$, UC, UBC and $UB_2$ pellets

In this study, various reactive sintering processes were carried out from uranium dioxide ($UO_2$) powder provided by AREVA (sample no. 65496), which contained 0.25 % by mass of uranium-235. According to the supplier's certificate of analysis, the $UO_2$ powder contains impurities of chromium (Cr), nickel (Ni), and iron (Fe) measured respectively at 4, 6, and 16 μg/g of U. The graphite used as a carbon source was acquired from Cerac with a purity of 99.5 % and a particle size of 44 μm. Hexagonal boron nitride (BN) was supplied by GoodFellow with a purity of 99.5 % and a particle size of 10 μm and boron carbide ($B_4C$) by Sigma-Aldrich with a purity of 98 % and particle sizes above 10 μm.

The chemical reactions involved in the process are as follows:

- $UO_{2(s)} + 4C_{(s)} \rightarrow UC_{2(s)} + 2CO_{(g)}$
- $UO_{2(s)} + 3C_{(s)} \rightarrow UC_{(s)} + 2CO_{(g)}$
- $UO_{2(s)} + 3C_{(s)} \rightarrow UC_{(s)} + 2CO_{(g)}$,
followed by $2UC_{(s)} + 2BN_{(s)} \rightarrow 2UBC_{(s)} + N_{2(g)}$
- $2UO_{2(s)} + B_4C_{(s)} + 3C_{(s)} \rightarrow 2UB_{2(s)} + 4CO_{(g)}$

In order to synthesise UC or $UC_2$ by reactive sintering, a quantity of 5 g in the mandatory stoichiometric ratios of $UO_2$ and graphite powders was prepared and placed in a 50 mL tungsten carbide (WC) grinding bowl. Three 10 mm diameter, nine 7 mm, and eighteen 3 mm WC grinding balls were added to the bowl. The mixture was milled using a Retsch PM 100 planetary mill. The use of multiple ball sizes enhances grinding efficiency by ensuring a more homogeneous powder mixture [10], [11]. Subsequently, 25 mL of isopropanol were added to the mixture. The grinding program was performed at 600 rpm for 20 minutes, with a reversal of rotation direction every 5 minutes. At the end of the program, the powder mixture was retrieved and exposed to ultrasound for 5 minutes using a Branson 3800 ultrasonic bath operating at 40 kHz. The mixture was then dried under a fume hood on a heated sand bath at 80 °C until all the solvent evaporated. The dried mixture was compressed to form five pellets of approximately 1 g each using a 13 mm diameter mould and uniaxially pressed, applying 220 MPa for 6 seconds. The pellets were then placed in a furnace similar to those used in target ion source system for radioactive beam production experiments, as described by Sundell and Ravn [12] and heated to a temperature of 1800 °C for 2 hours under secondary vacuum at $10^{-6}$ mbar to form UC or $UC_2$.

UBC pellets were obtained in two steps. Initially, UC pellets were synthesised in accordance with the previously described protocol. These pellets were then transferred to a glove box under an argon atmosphere to be manually ground in an agate mortar. The UC powders thus obtained were mixed in the appropriate stoichiometric proportions with BN, forming a batch of 2 g. The mixture was added to a beaker containing 50 mL of isopropanol and dispersed by ultrasound for 15 minutes. After dispersion, the solution was dried and then pressed as previously described. Due to their friability, the pellets could not be handled with tweezers and were carefully transferred into the furnace by gently sliding them off spatulas. Despite their delicate nature, all pellets were successfully processed without discarding any samples. The pellets were then heated in the same furnace as used for UC and $UC_2$ synthesis at 1800 °C for 2 hours under secondary vacuum.

The protocol established by Turner et al. [13], [14] allowed, after adaptation to our device, the synthesis of $UB_2$ pellets. Indeed, we built a furnace composed of two tantalum crucibles as described in Guillot et al. [15] to replicate as closely as possible the conditions used by Turner et al. It should be noted that the degassing caused during the synthesis severely damages the furnace. A 5 g sample of precursor powders in accordance with the required stoichiometric proportions was introduced into a 50 mL WC grinding bowl with the same set of balls in the planetary mill. A dry grinding was performed under the same conditions as previously described. Then, the mixed powder was retrieved and dispersed in 200 mL of isopropanol, with a brief application of ultrasound. The powder was retrieved, dried and compressed into five pellets of approximately 1 g using the same pressing process. These pellets were also heated to 1800 °C for 2 hours under secondary vacuum.

In order to control the percentage of open porosity generated during the synthesis, we heated the pellets in such a way that, during the temperature rise up to 1800 °C, the successive degassings corresponding to the chemical reactions did not degrade the vacuum beyond $10^{-4}$ mbar. This allowed us to limit the formation of excessive porosity due to gas emissions. Finally, we used the same holding time of 2 hours at high temperature (1800 °C), thus ensuring an identical sintering time for each sample in order to limit the dispersion of specific surface areas. The samples were all stored in elastic membrane boxes placed in a static vacuum chamber at $10^{-3}$ mbar to minimise oxidation.

### 2.2. Physicochemical characterization of the pellets

Following each synthesis, the samples were analysed by X-Ray Diffraction (XRD - Bruker D8 Advance) in the angular range between 20 and 90 $°2\theta_{Cu}$ as illustrated on the abscissa of Fig. 1. Phase identification was performed using the JCPDS-PDF-2 database, the different files used are mentioned in Fig. 1. A Rietveld refinement was conducted using the Maud software [16] to quantify the different phases in the 4-type synthesised samples.

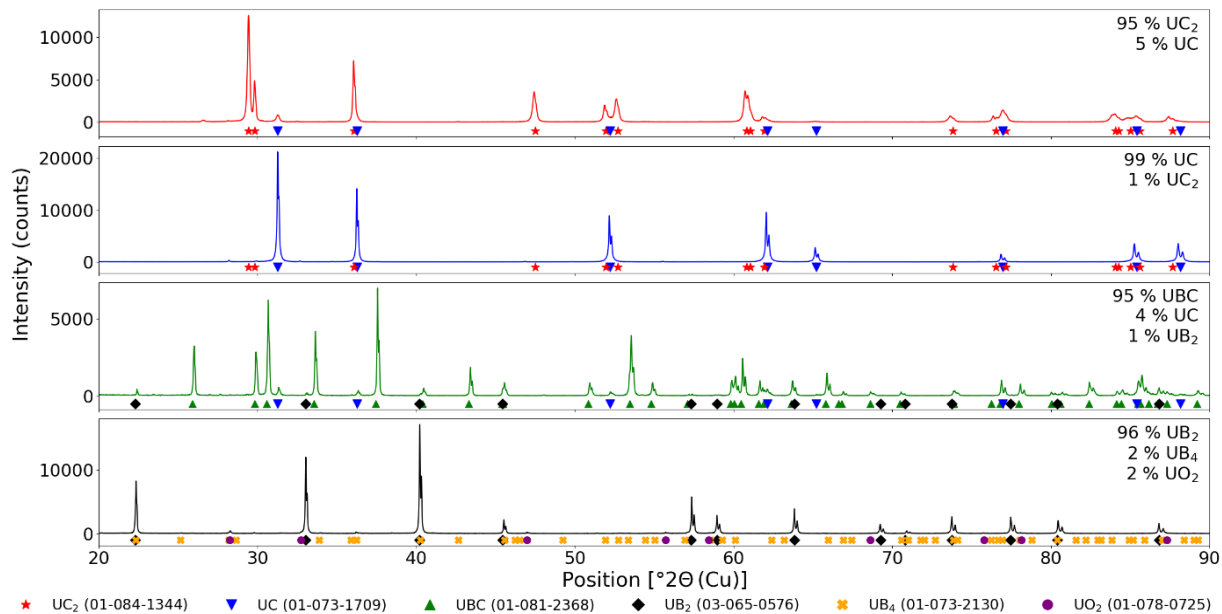

Fig. 1: Diffractograms of the pellets synthesised for the study, $UC_2$ in red, UC in blue, UBC in green and $UB_2$ in black. Symbols (star, inversed triangles, diamonds, …) are used to mark the expected position of Bragg lines for the various phases referred in the legend on bottom of the figure. For all the samples, the phase quantities from the Rietveld refinements are mentioned, the agreement factors were in the ranges: 10.3 % < $R_b$ < 15.1 %, 6.5 % < $R_{exp}$ < 8.0 %, 1.7 < $\chi^2$ < 2.8

Table 1: Summary of data obtained

| Target | Mass (g) | Diameter (mm) | Thickness (mm) | Apparent density (g/cm$^3$) | Porosity (%) Open | Porosity (%) Close | SSA (m$^2$/g) | Open pore size distribution (%) 0.1 - 10 µm | Open pore size distribution (%) 100 - 150 µm |
|---|---|---|---|---|---|---|---|---|---|
| $UC_2$ | 0.77 | 11.30 | 1.35 | 5.86 | 46 | 3 | 0.3965 | 88 | 12 |
| UC | 0.77 | 10.20 | 1.15 | 8.16 | 39 | 1 | 0.0763 | 94 | 6 |
| UBC | 0.92 | 12.76 | 1.02 | 6.93 | 42 | 1 | 0.0496 | 100 | 0 |
| $UB_2$ | 0.75 | 10.98 | 1.42 | 5.78 | 53 | 2 | 0.1032 | 80 | 20 |

In none of the syntheses was the pure compound obtained; however, the desired major phase was present in more than 95 % of the samples. For the remainder of the study, the sample will be considered and referred to by its major phase.

Helium pycnometry (Micromeritics ACCUPYC 1330) analysis allowed the apparent density of each sample to be measured, thus enabling the quantities of open and closed porosity formed by degassing to be determined. Mercury porosimetry (Micromeritics Autopore IV 9500) was used to quantify the diameter of open pores. The specific surface area (SSA) analysis by the Brunauer Emmett and Teller (BET) method was obtained with a Micromeritics ASAP 2020. The results of these analyses are presented in Table 1.

During these syntheses, we endeavoured to obtain pellets with an open porosity of approximately 45 % and a closed porosity of less than 3 %, as well as a similar open pore size distribution. These criteria were selected to eliminate the influence of porosity and thus permit the observation of the sole influence of the crystal during the release of fission products. The results of the SSA measurements on UC, UBC, and $UB_2$ pellets are comparable allowing a pertinent comparison of their release performance. However, the SSA of $UC_2$ is considerably higher, suggesting a finer porous structure increasing surface roughness, which may have a positive impact on the release. Scanning electron microscopy (Zeiss, SEM-FEG Sigma HD) images were obtained to account for the sample morphology. These images are presented in Fig. 2.

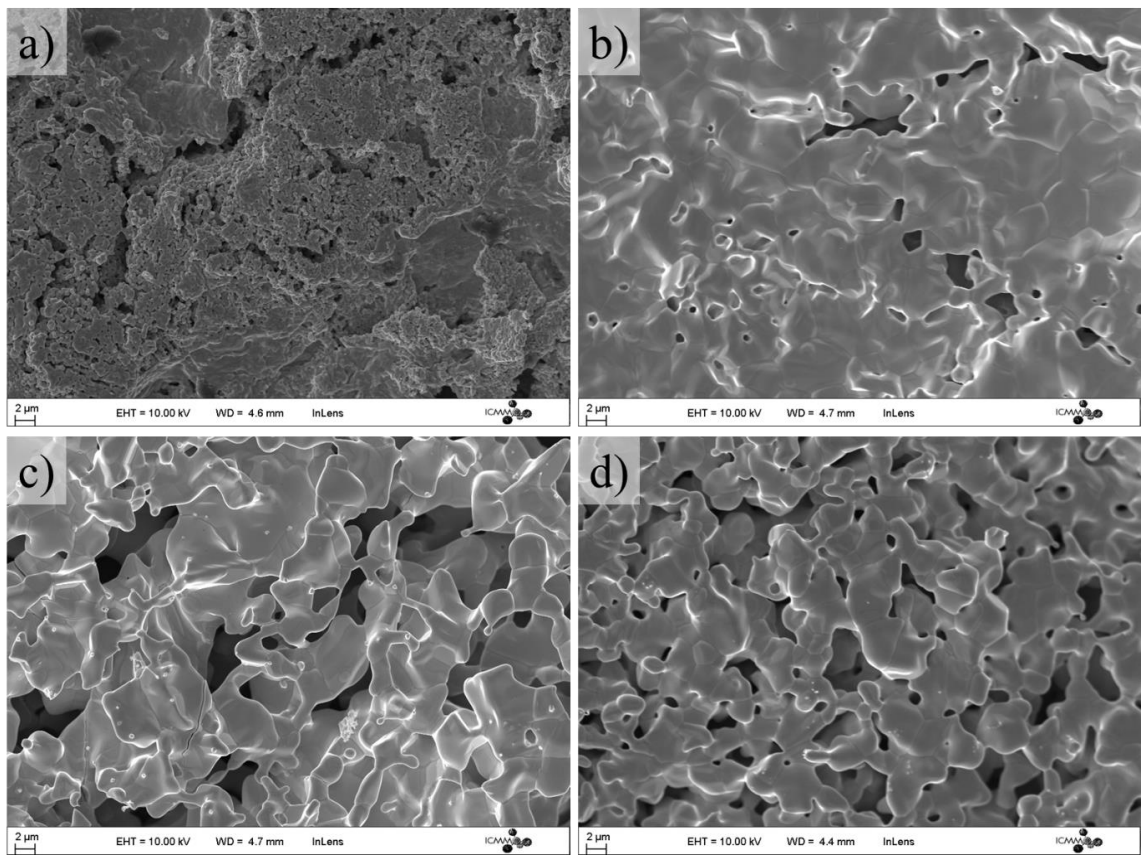

Fig. 2: SEM images of the synthesized uranium compounds: a) $UC_2$ sample exhibiting finer pores and increased surface roughness compared to other samples. b) UC sample showing micrometric grains ranging from 2 to 10 µm with a relatively smooth surface. c) UBC sample displaying a similar micrometric structure to UC, with grains between 2 and 10 µm. d) $UB_2$ sample also presenting micrometric grains comparable to UC and UBC.

The SEM results are consistent with those of the SSA measurements, indicating a similarity in the micrometric structure between the UC, UBC, and $UB_2$ samples, with an estimated grain size between 2 and 10 µm. The $UC_2$ sample (Fig. 2a) exhibits a distinct microstructure, comprising finer pores and increased surface roughness, which explains the higher SSA measurement observed for this sample.

2.3. Measurement of released fractions

The released fractions of eleven elements (Kr, Sr, Ru, Sn, Sb, Te, I, Cs, Ba, La, Ce) were measured off-line according to the procedure described in references [9], [15]. For each element, specific isotopes were identified by observing pure γ lines in the spectra corresponding to the energies of the γ transitions associated with their decay (Table 2). In contrast to our previous laboratory studies, where targets were heated to 1768 °C (the temperature corresponding to the melting point of platinum), we chose to work at a higher temperature, at 1789 °C in order to be in the same crystalline conditions as in an on-line experiment. Indeed, during on-line experiments, the target is heated to 2000 °C while irradiated with the electron beam. Thus, since according to the $UC_2$ phase diagram [17], the transition from the tetragonal configuration (α-$UC_2$) to the cubic configuration (β-$UC_2$) takes place above 1777 °C, this is a $UC_2$ crystal in its β-$UC_2$ configuration which is irradiated in on-line experiments. The extrapolation of the data obtained at 1800 °C to 2000 °C must be made with caution. Indeed, this 200°C difference between off-line and on-line experiments may be accompanied by a slight increase in the lattice parameter [18], which may favour diffusion. However, grain growth and intergranular sintering must also be taken into account, which may lead to a reduction in the release of fission products [19]. To ensure that the temperature limit of 1777 °C is exceeded, the sample heating temperature measurement has been improved and is now ensured by reading and recording data provided by an Omega type C thermocouple (T5R-010-12).

Table 2: Elements measured. The mass and the half-life of the isotopes used are indicated as well as the energy of the transitions signing the isotope decay and observed as pure γ lines in the spectra.

| Element | Z | A | $T_{1/2}$ | Energy (in keV) of the transitions observed as pure γ lines in the spectra |
|---|---|---|---|---|
| Kr | 36 | 88 | 2.84 h | 196.3, 898.0*, 1836.0*, 2195.8, 2392.1 |
| Sr | 38 | 91 | 9.63 h | 749.8, 1024.3 |
|    |    | 92 | 2.71 h | 1383.9 |
| Ru | 44 | 105 | 4.44 h | 469.4, 724.3 |
| Sn | 50 | 128 | 59.07 m | 482.3 |
| Sb | 51 | 129 | 4.40 h | 812.8 |
|    |    | 130 | 39.5 m | 330.9, 793.4 |
| Te | 52 | 132 | 3.204 d | 228.2 |
|    |    | 133m | 55.4 m | 334.2 |
|    |    | 134 | 41.8 m | 201.2, 277.9, 566.0 |
| I | 53 | 133 | 20.8 h | 529.9 |
|   |    | 134 | 52.6 m | 540.8, 595.4, 621.8, 847.0, 857.3, 884.1, 1072.5, 1136.2, 1613.8, 1806.8 |
|   |    | 135 | 6.57 h | 1131.5, 1260.4, 1678.0, 1791.1 |
| Cs | 55 | 138 | 33.41 m | 1435.9, 2218.0 |
| Ba | 56 | 139 | 83.06 m | 165.9 |
| La | 57 | 142 | 91.1 m | 641.3, 894.9, 1901.3, 2187.2, 2397.8, 2542.8 |
| Ce | 58 | 143 | 33.04 h | 293.3 |

*This transition belongs to the $^{88}$Rb→$^{88}$Sr decay. The $^{88}$Rb half-life is equal to 17.78 m; thus in view of the irradiation, waiting and counting times, the $^{88}$Rb nuclei that decay were not produced by photofission but come from the decay of $^{88}$Kr. Therefore, the behaviour of this γ line gives information about the Kr release.

The off-line experiment was conducted at the Tandem accelerator of the ALTO platform at the Irène Joliot-Curie laboratory and we will now briefly recall the major steps. In order to produce fission products, two pellets ($P_1$ and $P_2$) are irradiated with a 26 MeV deuteron beam with an electrical intensity of 20 nA for 20 minutes (corresponding to $1.5 \times 10^{14}$ particles). This duration is necessary to generate sufficient activity to be measured in the pellets, while still allowing them to be handled within the constraints of radioprotection, following ALARA principles. At the end of irradiation, an initial activity measurement of the two pellets is performed by γ spectrometry to determine the intensity ratio of the pure transitions characterizing the elements studied between the two pellets ($R = I_{P1}/I_{P2}$). The activities of the two pellets differed by 3 % to 10 %, with 10 % being the maximum observed difference. Subsequently, the $P_1$ pellet is subjected to a heating step at 1789 °C, maintained at this temperature to within ± 11 °C for 30 minutes. During the heating step, temperature ramp is applied to reach the temperature plateau within 10 minutes. These conditions (1789 °C for 30 minutes) were chosen to minimize the impact of the temperature ramp and to ensure a minimal heating time while retaining sufficient activity for measurement. Furthermore, 1789 °C is a temperature that allows us not to release the elements too quickly, thereby enabling the study of their release properties. After the heating, a new γ-spectrometry measurement is performed on the two pellets. The ratio of the intensities measured for the pure transitions characterising a given element between the two pellets, corrected for the previously defined ratio ($R$), allows the determination of the released fraction for this element:

$$RF(\%) = 100 \times \left(1 - \frac{I_{P1}}{R \times I_{P2}}\right) \qquad (4)$$

Therefore, the $RF$ are obtained for a temperature of 1789 °C and a heating dwell time of 30 min. It is important to note that between each stage (irradiation, γ measurement and heating) the samples are exposed to air and may undergo oxidation.

3. Results

The results of these measurements are presented in Fig. 3. In this figure, the experimental released fractions are represented by dots, and to guide the eyes by lines connecting them. Therefore, elements that could not be measured off-line are not extrapolated in these lines.

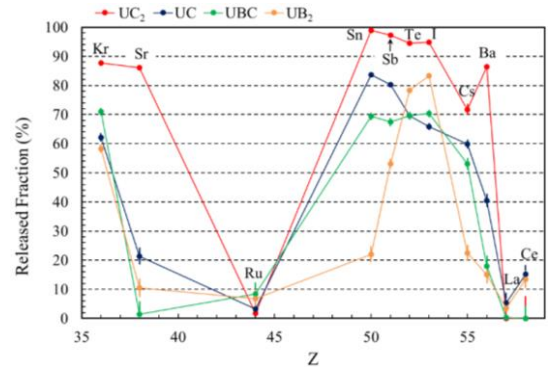

Fig. 3: Released fractions obtained for the UC$_2$, UC, UBC and UB$_2$ samples at 1789 ± 11 °C. Measurement uncertainties were calculated using the method presented in [9]. The largest contribution in the individual $I_{P1}/(R.I_{P2})$ error comes from the R error. The error adopted for RF is given by the maximum value between the individual errors and the deviations from the mean.

A first observation can be made: regardless of the sample, the elements Ru, La, and Ce are either not released or only in small quantities. It should be noted that Ru, a refractory element, has a very high melting point (2333 °C), so the heating temperature used in this experiment was insufficient to release it. With regard to La and Ce, although their melting points are lower (920 and 799 °C respectively) and their high boiling points at $10^{-5}$ mbar, 1307 °C and 1302 °C respectively, not prohibitive for our measurements, they are known to be difficult to release from uranium carbide targets even at high temperatures. They are preferably released as molecular compounds [20]. Consequently, the low release observed for these three elements can be attributed to the fact that between the irradiation step and the γ-spectrometry measurement with heating, the samples were exposed to air, which can induce oxidation and the formation of molecular compounds. Indeed, the work of Köster *et al.* [21] demonstrated the volatile nature of ruthenium as an oxide. As a consequence, Ru compounds probably escaped from the samples in the degassing we observed during the heating steps.

The second observation that emerges is that for the elements that are well released, the $UC_2$ sample appears to be the most efficient. It is important to note that this sample has a higher SSA, which has been demonstrated to be one of the parameters significantly affecting the release. Conversely, with the exception of iodine and tellurium, the $UB_2$ target exhibits the poorest release properties, which aligns with what Ravn *et al.* reported in reference [22] about the release of reaction products by borides. However, Kronenberg *et al.* nuance this by suggesting that $UB_{12}$ with its typical crystalline structure would be a favourable target from a release perspective [23].

Finally, it is notable that the order of release values observed for an element varies between different targets. This phenomenon will be discussed in greater detail in the subsequent section.

4. Discussion

Each of the four targets has a different crystallography: β-$UC_2$ (cubic), UBC (orthorhombic), UC (cubic), and $UB_2$ (hexagonal) with distinct packing fraction which is the part of a crystal's volume filled by its particles. Additionally, the 11 studied elements have various atomic radii, and except for Kr, the chemical environment of the elements can also play a role. All these factors can influence the release of elements and will be discussed in the following subsections.

4.1. Influence of crystal packing fraction and the size of the studied element on the release fraction

Fig. 4 presents the evolution of measured released fractions for the 11 studied elements as a function of crystal packing fraction. The packing fraction of the studied phases, calculated from the crystal lattice parameters and the atomic radii of the atoms contained in the lattice, is plotted on the abscissa. The lattice parameters used are for the β-$UC_2$ phase, the experimental values obtained in Chang's article [24], while those for the UBC, UC, and $UB_2$ phases are sourced from the JCPDS-PDF-2 identification cards. The atomic radii of U, B, and C used are the experimental values given in reference [25].

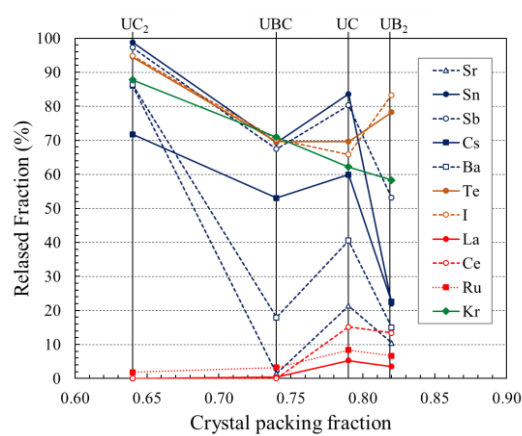

Fig. 4: Released fraction versus crystal packing fraction. The plot shows the measured released fractions (%) of 11 fission products from $UC_2$, UC, UBC, and $UB_2$ targets as a function of their crystal packing fractions. Each element is represented by a different symbol, and data points for each element are connected by lines for clarity. The figure demonstrates that Kr release increases linearly with decreasing packing fraction, while other elements show varying behaviours, suggesting that factors other than packing fraction influence their release.

Despite an apparent disorganisation, four groups of elements can be distinguished. The first group, formed by Kr, shows a linear evolution of the released fraction as a function of crystal packing fraction. This can be explained by the fact that, as a noble gas, Kr does not form chemical bonds with the crystal elements. In the second group are Ru, La, and Ce, the elements little or not released by the studied crystals. The third group includes Te and I, which alone present a notably higher released fraction by $UB_2$ than by UBC and UC, two crystals with lower packing fraction. Finally, for elements Sr, Sn, Sb, Cs, and Ba, which form the fourth group, the evolution of released fractions with packing fraction is very irregular. For these elements, release seems more influenced by the chemical environment of the crystal than by its packing fraction. Indeed, the presence of boron has a strong influence on the released fractions. Their values are observed to be lower when one or two boron atoms are present in the target crystal. With the exception of Sr, the released fractions by $UB_2$ are the lowest obtained for this group. The influence of the chemical properties of the targets will be discussed in the following section.

It is evident that crystal packing fraction is not the sole determining factor in the release of an element. Nevertheless, as this factor plays a role, it is of interest to examine the influence of the released element size. Fig. 5 presents the evolution of the measured released fractions for the four samples as a function of the released element atomic radius.

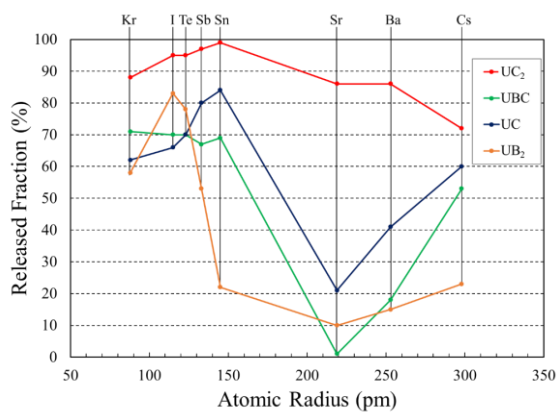

Fig. 5: Released fractions as a function of the atomic radius of the element released. Ru, La and Ce were not released by any of the four crystals studied and are not shown in the figure. This figure shows the relationship between the atomic radius of the released elements and their measured released fractions from the four target materials ($UC_2$, UC, UBC, $UB_2$). The graph helps analyse the influence of atomic size on the release efficiency, indicating how larger atomic radii may affect the mobility of elements within different crystal structures.

One would expect that the released fraction decreases as the atomic radius increases, with a more pronounced effect as the crystal packing fraction rises. Fig. 5 illustrates this phenomenon, demonstrating a falling off in the released fraction between the Sn and Sr elements, whose atomic radii are 150 and 220 pm, respectively. For $UC_2$, this decrease in the released fraction continues for larger atomic radii, whereas, at the opposite, an increase in the released fraction is observed between 220 and 300 pm for $UB_2$, and, with a stiffer slope, for UBC, UC that have a low packing fraction.

With regard to atomic radii, below 150 pm, an increase in the released fractions with the radius is observed for $UC_2$ and UC crystals, with very stable release values for UBC. However, for $UB_2$, an increase is observed between Kr and I, followed by a violent decrease for larger radii.

Such non-linear behaviours of the released fractions as a function of crystal packing fraction and the atomic radius of the element indicates that other chemical properties, influence the element's mobility within the crystal, and must be invoked to account for the release of different elements by these four crystals.

### 4.2. Influence of the various chemical properties of the elements on the release fraction

Many other chemical properties of the element, in addition to the crystal structure, can play an important role in the release process. For example, valence electrons, oxidation state (OS), ionisation energy, Pauling electronegativity and electron affinity are directly related to the formation of chemical bonds. The atomic and ionic radii can influence the mobility of the element in the crystal. Volatility, defined as the tendency of an element to vaporize at a given temperature, affects how readily an element can be released from the crystal lattice upon heating. Thermal stability refers to the ability of an element to resist decomposition or chemical change at high temperatures; elements with high thermal stability may remain in the solid phase and exhibit lower release fractions. These properties are considered in our study through the melting and boiling points of the elements. The values associated with these properties are summarised in Table 3 for all elements between iron (Fe) and dysprosium (Dy), i.e. those produced by the fission of $^{238}U$.

Table 3: Summary of physicochemical data used in the for Principal Component Analysis ([a][26], [b][27], [c][28], [d][29], [e][30], [f][25], [g][31]) performed in this work

| Atoms | Valence electrons [a] | OS (min) [a] | OS (max) [a] | Pauling electronegativity [a, b] | Electronic affinity (eV) [a, c, d] | Ionisation Energy (eV) [a] | Atomic radius (pm) [e, f] | Ionic radius** (pm) [a] | Boiling point at $10^{-5}$ mbar (°C) [g] | Melting point (°C) [a] | RF $UC_2$ (%) | RF UC (%) | RF UBC (%) | RF $UB_2$ (%) |
|---|---|---|---|---|---|---|---|---|---|---|---|---|---|---|
| Fe | 8 | 2 | 3 | 1.83 | 0.151 | 7.902 | 156 | 55$^{(+3)}$ | 1112 | 1538 | - | - | - | - |
| Co | 9 | 2 | 3 | 1.88 | 0.662 | 7.881 | 152 | 65$^{(+2)}$ | 1177 | 1495 | - | - | - | - |
| Ni | 10 | 2 | 3 | 1.91 | 1.157 | 7.640 | 149 | 69$^{(+2)}$ | 1167 | 1455 | - | - | - | - |
| Cu | 11 | 1 | 2 | 1.9 | 1.236 | 7.726 | 145 | 73$^{(+2)}$ | 932 | 1084.6 | - | - | - | - |
| Zn | 12 | 2 | 2 | 1.65 | -0.490 | 9.394 | 142 | 74$^{(+2)}$ | 209 | 419.53 | - | - | - | - |
| Ga | 3 | 3 | 3 | 1.81 | 0.301 | 5.999 | 136 | 62$^{(+3)}$ | 752 | 29.765 | - | - | - | - |
| Ge | 4 | 2 | 4 | 2.01 | 1.233 | 7.899 | 125 | 53$^{(+4)}$ | 1027 | 938.25 | - | - | - | - |
| As | 5 | -3 | 5 | 2.01 | 0.804 | 9.789 | 114 | 58$^{(+3)}$ | 207 | 817 | - | - | - | - |
| Se | 6 | -2 | 6 | 2.55 | 2.021 | 9.752 | 103 | 50$^{(+4)}$ | 134 | 220.8 | - | - | - | - |
| Br | 7 | -1 | 5 | 2.96 | 3.364 | 11.814 | 94 | 196$^{(-1)}$ | -140 | -7.2 | - | - | - | - |
| Kr | 8 | 0 | 0 | 3 | -2.41 | 14.000 | 88 | - | -229.7* | -157.37 | 88 | 62 | 71 | 58 |
| Rb | 1 | 1 | 1 | 0.82 | 0.486 | 4.177 | 265 | 152$^{(+1)}$ | 62 | 39.3 | - | - | - | - |
| Sr | 2 | 2 | 2 | 0.95 | 0.052 | 5.685 | 219 | 118$^{(+2)}$ | 355 | 777 | 86 | 21 | 1 | 10 |
| Y | 3 | 3 | 3 | 1.22 | 0.307 | 6.217 | 212 | 90$^{(+3)}$ | 1227 | 1522 | - | - | - | - |
| Zr | 4 | 4 | 4 | 1.33 | 0.426 | 6.634 | 206 | 72$^{(+4)}$ | 1877 | 2854 | - | - | - | - |
| Nb | 5 | 3 | 5 | 1.6 | 0.917 | 6.759 | 198 | 64$^{(+5)}$ | 2117 | 2477 | - | - | - | - |
| Mo | 6 | 6 | 6 | 2.16 | 0.746 | 7.092 | 190 | 59$^{(+6)}$ | 1937 | 2622 | - | - | - | - |
| Tc | 7 | 4 | 7 | 2.1 | 0.550 | 7.280 | 183 | 65$^{(+4)}$ | 1967 | 2157 | - | - | - | - |
| Ru | 8 | 3 | 3 | 2.2 | 1.046 | 7.361 | 178 | 68$^{(+3)}$ | 1857 | 2333 | 2 | 3 | 8 | 7 |
| Rh | 9 | 3 | 3 | 2.28 | 1.143 | 7.459 | 173 | 67$^{(+3)}$ | 1582 | 2963 | - | - | - | - |
| Pd | 10 | 2 | 3 | 2.2 | 0.562 | 8.337 | 169 | 86$^{(+2)}$ | 1097 | 1554.8 | - | - | - | - |
| Ag | 11 | 1 | 1 | 1.93 | 1.304 | 7.576 | 165 | 115$^{(+1)}$ | 747 | 961.78 | - | - | - | - |
| Cd | 12 | 2 | 2 | 1.69 | -0.330 | 8.994 | 161 | 95$^{(+2)}$ | 145 | 321.07 | - | - | - | - |
| In | 3 | 3 | 3 | 1.78 | 0.404 | 5.786 | 156 | 80$^{(+3)}$ | 663 | 156.60 | - | - | - | - |
| Sn | 4 | 2 | 4 | 1.96 | 1.112 | 7.344 | 145 | 69$^{(+4)}$ | 897 | 231.93 | 99 | 84 | 69 | 22 |
| Sb | 5 | -3 | 5 | 2.05 | 1.047 | 8.608 | 133 | 76$^{(+3)}$ | 385 | 630.63 | 97 | 80 | 67 | 53 |
| Te | 6 | -2 | 6 | 2.1 | 1.971 | 9.010 | 123 | 97$^{(+4)}$ | 255 | 449.51 | 95 | 70 | 70 | 78 |
| I | 7 | -1 | 7 | 2.66 | 3.059 | 10.451 | 115 | 220$^{(-1)}$ | -67 | 113.7 | 95 | 66 | 70 | 83 |
| Xe | 8 | 0 | 0 | 2.6 | -1.76 | 12.130 | 108 | - | -218* | -111.75 | - | - | - | - |
| Cs | 1 | 1 | 1 | 0.79 | 0.472 | 3.894 | 298 | 167$^{(+1)}$ | 46 | 28.5 | 72 | 60 | 53 | 23 |
| Ba | 2 | 2 | 2 | 0.89 | 0.145 | 5.212 | 253 | 135$^{(+2)}$ | 453 | 727 | 86 | 41 | 18 | 15 |
| La | 3 | 3 | 3 | 1.1 | 0.558 | 5.577 | 195 | 103$^{(+3)}$ | 1307 | 920 | 0 | 5 | 0 | 4 |
| Ce | 4 | 3 | 4 | 1.12 | 0.628 | 5.539 | 185 | 101$^{(+3)}$ | 1302 | 799 | 0 | 15 | 0 | 13 |
| Pr | 5 | 3 | 3 | 1.13 | 0.109 | 5.473 | 247 | 99$^{(+3)}$ | 1092 | 931 | - | - | - | - |
| Nd | 6 | 3 | 3 | 1.14 | 0.097 | 5.525 | 206 | 98$^{(+3)}$ | 967 | 1016 | - | - | - | - |
| Pm | 7 | 3 | 3 | - | 0.129 | 5.582 | 205 | 97$^{(+3)}$ | 847 | 1042 | - | - | - | - |
| Sm | 8 | 2 | 3 | 1.17 | 0.162 | 5.644 | 238 | 96$^{(+3)}$ | 511 | 1072 | - | - | - | - |
| Eu | 9 | 2 | 3 | - | 0.864 | 5.670 | 231 | 95$^{(+3)}$ | 409 | 822 | - | - | - | - |
| Gd | 10 | 3 | 3 | 1.2 | 0.137 | 6.150 | 233 | 94$^{(+3)}$ | 1177 | 1313 | - | - | - | - |
| Tb | 11 | 3 | 3 | - | 0.131 | 5.864 | 225 | 92$^{(+3)}$ | 1137 | 1359 | - | - | - | - |
| Dy | 12 | 3 | 3 | 1.22 | 0.352 | 5.939 | 228 | 91$^{(+3)}$ | 812 | 1412 | - | - | - | - |

* extrapolated from ref. [26]

** ionic radius corresponding to the most probable oxidation state indicated by superscript

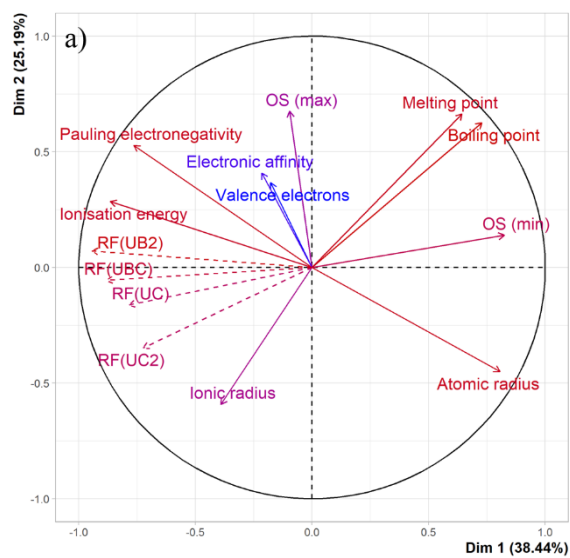

| | Dim.1 | | Dim.2 | |
|---|---|---|---|---|
| | correlation | p-value | correlation | p-value |
| Valence electrons | -0.1770 | 2.68E-01 | 0.3659 | 1.86E-02 |
| OS (min) | 0.8247 | 3.38E-11 | 0.1397 | 3.84E-01 |
| OS (max) | -0.0953 | 5.53E-01 | 0.6749 | 1.30E-06 |
| Pauling electronegativity | -0.7633 | 6.50E-09 | 0.5273 | 3.97E-04 |
| Atomic radius | 0.8062 | 2.01E-10 | -0.4504 | 3.13E-03 |
| Ionic radius | -0.3903 | 1.16E-02 | -0.5923 | 4.51E-05 |
| Boiling point | 0.7287 | 6.61E-08 | 0.6250 | 1.25E-05 |
| Melting point | 0.6431 | 5.77E-06 | 0.6634 | 2.28E-06 |
| Electronic affinity | -0.2155 | 1.76E-01 | 0.4080 | 8.10E-03 |
| Ionisation energy | -0.8645 | 3.24E-13 | 0.2844 | 7.15E-02 |
| RF(UC$_2$) | -0.7238 | 8.91E-08 | -0.3483 | 2.56E-02 |
| RF(UC) | -0.7853 | 1.21E-09 | -0.1622 | 3.11E-01 |
| RF(UBC) | -0.8815 | 2.79E-14 | -0.0551 | 7.32E-01 |
| RF(UB$_2$) | -0.9440 | 2.24E-20 | 0.0712 | 6.58E-01 |

Fig. 6: PCA variable results on physicochemical properties and release fractions of the elements studied. a) Graph of variables: Variable correlation circle displaying the projection of active variables (physicochemical properties such as electronegativity, atomic radius, melting point, etc.) and supplementary variables (released fractions for UC$_2$, UC, UBC, and UB$_2$) onto the first two principal components (Dim. 1 and Dim. 2). The angles and lengths of the vectors represent the correlations and contributions of each variable to the principal components, helping to identify relationships between variables and their influence on element release. b) Table summarizing the correlation coefficients and p-values of the active and supplementary variables with respect to the first factorial plane (Dim. 1 and Dim. 2). This information indicates how well each variable is represented in the PCA and its significance in explaining the variance in the data set.

In order to determine if there were any correlations between the physicochemical properties of the elements and their release characteristics, a Principal Component Analysis (PCA) was used [32], [33]. PCA is a multivariate statistical method that allows the information contained in a data set to be analysed with the aim of identifying linear correlations between variables and to synthesise this information by reducing the number of variables. To study the impact of variables expressed in different units, the correlation matrix must be computed using the standardised (i.e. centred and scaled) data. Diagonalising this matrix gives access to the principal components, which are the unit eigenvectors of the correlation matrix. PCA provides a graphical representation of the data set in a two-dimensional space defined by the first two principal components (referred to as Dim.1 and Dim.2 in the following graphs).

The 41 elements presented in Table 3 contributed to the PCA through their 10 physicochemical properties. For these 10 variables, missing data were imputed using a 2-dimensional PCA model [34]. The release variables, considered as supplementary variables, did not contribute to the construction of the factorial axes. If strong correlations are observed between the physicochemical properties of the elements and their release properties, it will be possible to propose which elements are likely to be released by the four crystals studied. Figure 6 shows the PCA results obtained on the variables.

The inertia projected on this first factorial plane corresponds to 63.63 % of the total data-set inertia, with 38.44 % in dimension 1 and 25.19 % in dimension 2. This means that 63.63 % of the total point-cloud variability is represented in this plane. This value is significantly higher than the 95 % quantile inertia value for 10 independent variables and 41 individuals, which is 37.9 % [32], [33]. Therefore, the variability explained by this plane is highly significant. Consequently, the analysis can be restricted to this first principal plane, from which we will extract the physical meaning in the following paragraphs.

An important evaluator in PCA is the cos$^2$, which measures the quality of the variable representation. Its value is indicated by a colour code. Fig. 6a shows that most variables, including the supplementary release variables, are well described in the plane (Dim. 1, Dim. 2) as they have a cos$^2$ greater than 0.6. The variables "Valence", "OS (max)", "Electronic affinity", and "Ionic radius", with cos$^2$ values of 0.17, 0.46, 0.21, and 0.50, respectively, are poorly represented in this plane.

The correlation and critical probability (p-value) values obtained between the two main PCA axes and the 14 variables are presented in the table in Figure 6b. The correlation value indicates the strength (magnitude) and direction (sign) of the linear relationship between each variable and the dimensions (Dim. 1 or Dim. 2), and the p-value evaluates the statistical significance of the observed correlation. Thus, a p-value below 0.05 indicates that the observed correlation is not random but statistically significant. The analysis of the correlation values, p-value, and $\cos^2$ obtained between the active variables and Dim. 1 and 2 allowed us to extract the physical meaning of these two principal components.

The first principal component (Dim. 1) is negatively correlated with Pauling electronegativity and ionisation energy. It is also positively correlated with the minimal oxidation state, atomic radius, melting point, and boiling point at $10^{-5}$ mbar. Therefore, Dim. 1 appears to be associated with the volatility of the elements and their reactivity within the crystal. The second principal component (Dim. 2) is positively correlated with the melting point and boiling point at $10^{-5}$ mbar. Dim. 2 thus appears to represent the thermal stability aspects of the elements.

The vectors representing the observed released fractions for the four samples $UC_2$, UC, $UBC$, and $UB_2$, supplementary variables that did not contribute to the construction of the main axes, appear to be negatively and very significantly correlated with Dim. 1. This indicates that the release processes are strongly associated with the variables defining this dimension. The two principal components capture a combination of physical and chemical properties that influence an element's ability to diffuse and thus to be released from the solid in which it diffuses. Elements with a high positive minimal oxidation state will tend to form bonds in the crystals studied; their mobility will be hindered if they have a large atomic radius and/or high melting and boiling points, opposing their release.

PCA not only examines correlations between variables but also highlights similarities between individuals. Fig. 7 shows the representation of the individuals (the 41 elements of our study) in the first factorial plane. The elements for which released fractions were measured are indicated by a star.

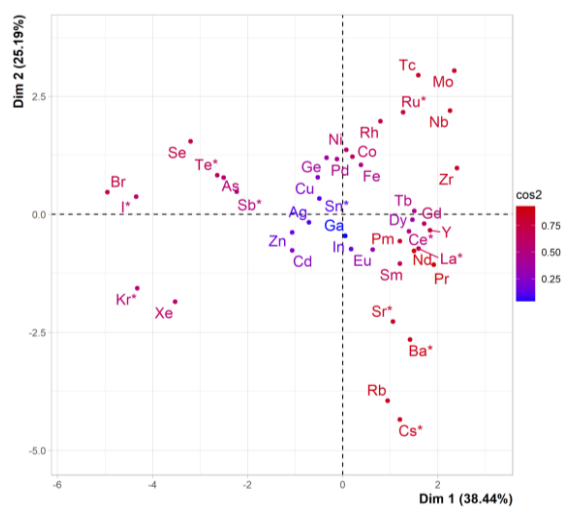

Fig. 7: Projection of the 41 elements onto the first factorial plane (Dim. 1 and Dim. 2) from the PCA analysis. Each point represents an element, positioned according to its physicochemical properties as determined by the PCA. Elements for which released fractions were measured (indicated by stars) are highlighted. The figure illustrates groupings of elements with similar properties, allowing for the identification of correlations between their physicochemical characteristics and release behaviours. Elements close to each other in this plane are expected to have similar release properties in the four crystals studied.

As with the variables, the $\cos^2$ indicates the quality of the element representation in the plane (Dim. 1, Dim. 2). The 13 elements with a $\cos^2$ below 0.5 (Fe, Cu, Zn, Ga, Ge, Pd, Ag, Cd, In, Sn, Tb, Eu and Dy) are poorly represented in this plane. These elements are all quite close to the origin of the graph, indicating that they have relatively low coordinates on Dim. 1 and 2 and are thus not strongly linked to the variables describing these two dimensions. Among them are five elements (Ga, Ag, Cd, In, and Sn), whose release properties by $UC_x$ targets are primarily governed by the effusion process rather than diffusion as shown in references [5], [35].

For the remaining 28 elements, the representation in the first factorial plane is relevant. They are grouped according to their "chemical family". Thus, in the top right of Fig. 7, elements with high coordinates on both Dim. 1 and Dim. 2, are characterized by high melting and boiling points at $10^{-5}$ mbar (Dim. 2) but also by high values of minimal oxidation state (Dim. 1). These are refractory elements (Zr, Nb, Mo, Tc, Ru, and Rh). Closer to the horizontal axis but still on the right side of the figure are the lanthanides, which also present a relatively high minimal oxidation state (2 or 3) but lower temperatures (melting and boiling). At the bottom right are the alkaline-earth and alkali metals with a high negative coordinate on Dim. 2 and a positive coordinate on Dim. 1. This is due to their low characteristic temperatures and large atomic radius. At the bottom of Fig. 7, on the left side, are the elements associated with negative coordinates on Dim. 1 and Dim. 2, corresponding to the smallest atomic radii and lowest melting points: noble gases (Xe and Kr). In the same left side of the plot, but with positive coordinates on Dim. 2, are elements characterized by negative minimal oxidation states. These include the halogens (Br, I), metalloids (As, Te, Sb), and a poor metal (Se).

The position of the elements in the plane (Dim. 1, Dim. 2) thus reflects well the physical and chemical properties captured by dimensions 1 and 2 of the PCA. The analysis of the variables showed that the release properties in the $UC_2$, UC, UBC, and $UB_2$ crystals were well described in the first factorial plane and strongly negatively correlated with Dim. 1. Furthermore, when for elements close in the first factorial plane, the release fractions by the four studied crystals are known, it is observed that the similarity in their chemical properties is accompanied by a resemblance in their release properties. For example, Te and Sb, two metalloids, are close on Fig. 7, and their release fractions by three of the crystals are very similar (for both: ~95 % by $UC_2$, ~75 % by UC, ~70 % by UBC) and differ only for $UB_2$ (53% for Sb and 78% for Te). Sr and Ba, the two alkaline earth metals, are released very well by $UC_2$ but much less by the other three crystals. It can therefore be assumed that when two elements are close to each other in the plane (Dim.1, Dim. 2), they are close not only in terms of their physicochemical properties, but also in terms of their release properties, which will be very similar in the four crystals studied. Br is close to I on Fig. 7, and both are halogens, Br is expected to be released similarly to I, especially by $UB_2$. Similarly, Rb, being close to another alkali metal, Cs, should be well released by $UC_2$, less well by UC and UBC, and very little by $UB_2$. Xe, a noble gas, and As, a metalloid, are close in Fig. 7 to two other elements of the same type, Kr and Sb, respectively, so they could have similar release properties in $UC_2$, UC, UBC, and $UB_2$. Se, which belongs to the same group 16 (IUPAC group) as Te, could show similar releases to those of Te: very good for $UC_2$ and a little less good for the other three crystals. In Fig. 7, the lanthanides are close to La and Ce, for which the release fractions have been measured, so they are also expected to be released in small amounts. Finally, the elements surrounding Ru are all refractory, and should not be released by any of the four crystals studied.

Our results and interpretations are consistent with previous studies that have investigated the release of fission products from uranium targets mainly in carbide form. Carraz et al. [36] and Ravn et al. [22] showed that $UC_x$ targets have a high efficiency for the rapid release of volatile elements such as Kr and I at high temperatures. Furthermore, Corradetti et al. [37] showed that diffusion and effusion are strongly influenced by the target temperature. In the study by Hoff et al. [38], lanthanides show an ability to diffuse within the crystalline structure of target materials, but their effusion towards the surface to pass into the gas phase remains very limited. The authors suggest that this limitation is often due to their low volatility and high affinity with the target materials, which prevents them from desorbing efficiently. Their observation seems to be consistent with the PCA interpretation of our results. To improve their release, Hoff et al. studied the introduction of reactive gases, such as $CF_4$ or $BF_3$. This method has been tested to improve release by forming volatile complexes, but some elements, such as Zr, still show resistance to release even with the addition of gas. Other studies, such as those by Köster et al. [21] and Kronenberg et al. [23], have also shown that refractory elements, such as Ru, require specific experimental conditions, suggesting the use of oxide targets like $UO_2$ and $ThO_2$. In 2022, Ballof et al. [39] highlighted the influence of volatility and thermal stability on release efficiency, emphasising the importance of vapour pressure and sublimation temperatures on the extraction efficiency of refractory elements such as Mo. They propose to use the kinetic energy of the fission recoil to move the fission products to the outside of the target, where they can react with carbon monoxide (CO) to form volatile carbonyl complexes. This approach avoids the need for high temperatures to extract certain refractory elements that are difficult to release.

5. Conclusion

In this study, four uranium compounds ($UC_2$, UC, UBC, and $UB_2$) were synthesised to investigate the release behaviour of 11 fission elements. The analysis of the experimental results showed that crystal packing fraction and atomic size affect element release. Krypton, a noble gas, shows a linear relationship between its released fraction and crystal packing fraction. In contrast, the release of other elements is influenced in a complex manner by the chemical environment and atomic size. Other properties beyond packing fraction and atomic size also play a significant role in the release of the elements studied.

A Principal Component Analysis (PCA) was performed on the physicochemical properties of all the elements produced by fission (from iron to dysprosium), considered as active variables, and on the released fractions, considered as supplementary variables. The release variables are well represented in the first factorial plane of the PCA and show strong correlations with electronegativity, ionisation energy, oxidation states, thermal stability (melting and boiling points), and atomic radius.

PCA revealed that chemical properties control the mobility and reactivity of elements. For alkali and alkaline earth metals, atomic size limits their mobility, suggesting that a crystal matrix with low packing fraction will favour their diffusion. Lanthanides, compared to the previous elements, are more likely to form chemical bonds, limiting their escape from the crystal. To facilitate their release, the introduction of fluorine has proven effective, forming molecular beams. Elements such as zirconium, niobium, molybdenum, technetium, ruthenium, and rhodium are characterised by very high melting and boiling points, which indicate their refractory nature and explain their stability at high temperature. This characteristic makes their extraction complex at the typical temperatures used in targets, i.e. 2000 °C. Therefore, exploring alternative methods, such as using uranium oxycarbide or uranium oxide targets, could improve the release of these elements without degrading the target material.

Finally, this study has shown that even if some elements are not directly observable during off-line experiments, their behaviour can likely be inferred from the releases measured for the 11 accessible elements. For example, the behaviour of bromine can be deduced from that of iodine, as can that of xenon from krypton, rubidium from caesium, lanthanides from lanthanum and cerium, and finally, niobium, molybdenum, technetium, and rhodium from ruthenium.

Acknowledgments:

The authors would like to express their gratitude to the management of IJCLab and the scientific coordinator for providing beam time at the ALTO facility. We also extend our sincere thanks to the ALTO team for their availability and assistance in preparing this experiment, as well as to the SPR team for ensuring radiation protection throughout the study.